# 3D MODEL RETRIEVAL USING GLOBAL AND LOCAL RADIAL DISTANCES

*Bo Li, Henry Johan*

School of Computer Engineering
Nanyang Technological University
Singapore
E-mail: libo0002@ntu.edu.sg, henryjohan@ntu.edu.sg

## ABSTRACT

3D model retrieval techniques can be classified as histogram-based, view-based and graph-based approaches. We propose a hybrid shape descriptor which combines the global and local radial distance features by utilizing the histogram-based and view-based approaches respectively. We define an area-weighted global radial distance with respect to the center of the bounding sphere of the model and encode its distribution into a 2D histogram as the global radial distance shape descriptor. We then uniformly divide the bounding cube of a 3D model into a set of small cubes and define their centers as local centers. Then, we compute the local radial distance of a point based on the nearest local center. By sparsely sampling a set of views and encoding the local radial distance feature on the rendered views by color coding, we extract the local radial distance shape descriptor. Based on these two shape descriptors, we develop a hybrid radial distance shape descriptor for 3D model retrieval. Experiment results show that our hybrid shape descriptor outperforms several typical histogram-based and view-based approaches.

## 1. INTRODUCTION

3D model retrieval is now recognized as one of the important research areas in computer graphics and multimedia. The fundamental part of a 3D model retrieval algorithm is the construction of a shape descriptor to represent and compare different models. In recent years, there have been many studies in proposing distinctive shape descriptors. They can be classified into three groups, which are histogram-based [1, 21, 14], view-based [4, 23, 20] and graph-based [8, 25].

It is well known that histogram-based approach is difficult to achieve high accuracy but it can extract shape features efficiently, and view-based approach can achieve better retrieval performance. Generally, it is difficult for a single type of shape descriptor to perform well for all kinds of models and usually a hybrid shape descriptor can perform better.

Motivated by the above facts, we propose a hybrid radial distance shape descriptor by fusing the histogram-based and view-based approaches. We represent global radial distance feature using a histogram-based descriptor and local radial distance feature using a view-based approach. Our local radial distance feature can already achieve a satisfying retrieval performance. Adding the global radial distance feature improves the performance further without adding much computation load because of an efficient histogram-based feature extraction process. Based on our experiments on two standard databases, we found our hybrid approach can outperform typical histogram-based approach, such as shape distribution [21, 22] and shape histogram [1] and view-based approach such as LightField descriptor [4]. The speed of our algorithm is also satisfying.

The rest of the paper is organized as follows. We discuss the related work in Section 2. Section 3 introduces the global and local radial distance features. In Section 4, we describe the 3D model retrieval algorithm based on the proposed hybrid radial distance shape descriptor. The performance of our approach is verified and compared using the PSB [24] and NTU [4] databases, which are detailed in Section 5. We conclude the work and list future topics in Section 6.

## 2. RELATED WORK

Natraj et al. [19] and Tangelder et al. [26] classified and compared different 3D model retrieval techniques in their respective surveys. In this section, we mainly review the related work in the histogram-based and view-based techniques. Histogram-based techniques use the distribution of features extracted based on geometric elements, such as vertices and faces to represent a 3D model. Both 1D and 2D histograms are often used. Shape distribution [21, 22], shape histogram [1] and 3D shape context [14, 18, 9, 6] are three typical histogram-based techniques.

Shape distribution [21, 22] focus on geometric shape function that measures the distance between two random points on the surface of the model. These probability distributions of the models are compared during the retrieval process.

Shape histogram [1] computes the distance from the center of mass and spherical angle for each surface point. Then it encodes the distance distribution into a histogram, whose bins are formed according to three types of 3D space partitioning methods: Shells (only use distance), Sectors (only use spherical angle) and Spider Web (use both).

3D shape context [14, 18, 6, 9] is based on the idea of 2D shape context [3], which is a log-polar histogram and defines the relative distribution of other points with respect to a point. The generalized shape context [18] directly extends the idea from 2D to 3D. There are three other forms of 3D shape context based on three different 3D space partitioning, named Shell, Sector and Spider. Shell model divides the 3D space into a set of concentric spheres [6]. Sector model divides the 3D space in the spherical angle space and partitions it into several sectors, and this makes the 3D cylindrical shape context [9]. Spider web model combines both [14].

Ben-Chen and Gotsman [2] proposed a new 3D shape descriptor named conformal factor which depicts the amount of local work involved to transform a model into a sphere. Then, they use the histogram of the conformal factor to depict the feature of a 3D model and use L1 distance to measure the difference between two feature histograms. Spherical harmonics [12], spherical wavelet [16] and moments-based approaches [5] can be also regarded as extended histogram-based techniques.

In the view-based techniques, features are extracted based on the rendered view images. The visual similarities between the rendered view images of different models are compared with each other to measure the difference of the models. LightField [4] is a famous view-based shape descriptor. It defines the distance of two models as the minimal distance between their 10 corresponding silhouette views. To measure the difference of two silhouette views, it uses a hybrid image metric [29] which integrates the Zernike moments descriptor and Fourier descriptor. Multiple view descriptor [10] first aligns the model with Principle Component Analysis (PCA) [11] and then classifies models by comparing the primary, secondary and tertiary views defined by the principle axes.

Salient local visual feature-based retrieval method [23] first renders a set of depth view images for a 3D model and then extracts the multi-scale local features of these views using Scale Invariant Feature Transform (SIFT) [17], which is invariant to translation, scaling and rotation. Finally, it fuses all these local features into a histogram using Bag-Of-Features (BoF) approach, which accumulates the visual words (extends from the bag-of-words in text retrieval) of multiple views into a single histogram to represent the feature of a 3D model.

Recently, Petros et al. [20] proposed another view-based descriptor named Compact Multi-View Descriptor (CMVD), which supports multimodal retrieval such as sketches, images and models-based retrieval. It utilizes both binary and depth view images of models.

## 3. GLOBAL AND LOCAL RADIAL DISTANCE FEATURES EXTRACTION

### 3.1 Overview of Feature Extraction

We use global and local radial distance features to represent a 3D model. The feature extraction procedure consists of two steps: first normalize the 3D model and then compute the two features. For the normalization part, we first compute the bounding sphere of the 3D model based on the algorithm in [28] and an efficient implementation of the algorithm developed in [7]. Then, we translate the model so that the center of the bounding sphere coincides with the origin of the coordinate system and then scale the model to make the radius of the bounding sphere to be 1.0. Next, we need to align the 3D models. We can perform an approximate alignment based on Principle Component Analysis (PCA) [11], continuous PCA [27], or other alignment algorithms. However, to focus on the descriptor itself and to test its optimal performance, in our experiments, we manually front-pose aligned the models. For the second step of feature extraction, we present the details in Section 3.2 and Section 3.3 respectively.

### 3.2 Global Radial Distance Feature Extraction

Based on the idea of shape histogram [1], we define a global radial distance descriptor. Global radial distance descriptor captures the geometric distribution of the surface of a 3D model. Since the center of its bounding sphere coincides with the origin after the coordinate normalization, the global radial distance of one point $p$ on the surface of a 3D model is just the magnitude of the vector from the origin to $p$. To compute the descriptor of a 3D model, we uniformly divide the spherical angle space of the bounding sphere into a set of bins and in each bin we store the average radial distance of all the points in the bin as the 3D radial distance value. For a vector **v** originated from the origin to a point $p$, we assume the angle between **v** and $y/x$ axis is $\varphi/\theta$ ($\varphi \in [0, 180]$, $\theta \in [0, 360)$) respectively. The intervals for dividing the $\varphi$ and $\theta$ angle spaces are $\Delta\varphi$ and $\Delta\theta$, then we define each bin as follows,

$$bin(i,j) = \{(\varphi,\theta) | i*\Delta\varphi \leq \varphi < (i+1)*\Delta\varphi, j*\Delta\theta \leq \theta < (j+1)*\Delta\theta, 0 \leq i < \frac{180}{\Delta\varphi}, 0 \leq j < \frac{360}{\Delta\theta}\}. \quad (1)$$

Assume a 3D model $H$ consists of $n$ vertices $\mathbf{V}=\{v_1,v_2,\ldots,v_n\}$, and $m$ faces $\mathbf{F}=\{f_1,f_2,\ldots,f_m\}$. The centers of $\mathbf{F}$ are $\mathbf{C}=\{c_1,c_2,\ldots,c_m\}$ and the areas of $\mathbf{F}$ are $\mathbf{A}=\{a_1,a_2,\ldots,a_m\}$. Since the numbers of vertices in similar 3D models may be drastically different, we propose an area-weighted radial distance shape descriptor,

$$H(i,j) = \frac{\sum_{c_k \in bin(i,j)} \|c_k\| * a_k}{\sum_{c_k \in bin(i,j)} a_k}. \quad (2)$$

It weights the radial distances of all the face centers in the bin by their respective face areas.

Based on our experiments, we found that choosing the centers of faces rather than the vertices directly can achieve better retrieval performances in terms of precision-recall plots. We tested segmenting the spherical angle space of the bounding sphere with different intervals: $\Delta\varphi=\Delta\theta=45°$, $\Delta\varphi=\Delta\theta=30°$, $\Delta\varphi=\Delta\theta=15°$, $\Delta\varphi=\Delta\theta=10°$ and found that fixing the steps to be 30° can achieve the best results. Therefore, in our experiments, we choose $\Delta\varphi=\Delta\theta=30°$, and thus divide the bounding sphere into 72 bins and the global radial distance descriptor is a 6×12 matrix. Examples of the global radial distance feature are shown in Figure 1. We can see that similar models have similar global radial distance descriptors and the radial distance distributions of different models are often distinctively different. It is also very fast to compute these global radial distance features.

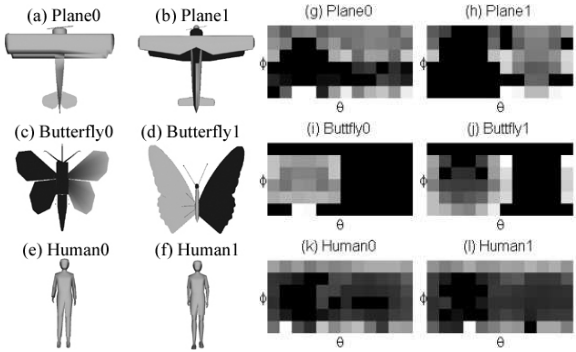

Fig. 1: Examples of global radial distance features.

### 3.3 Local Radial Distance Feature Extraction

Global radial distance only captures the global statistical property in the meaning of the radial distances with respect to the center of the bounding sphere of a 3D model. The main advantage of this approach is its high efficiency in terms of feature extraction time. However, the retrieval performance of this feature alone is not satisfying. To enhance retrieval performance, we combine it with a local radial distance feature.

Local radial distance is computed based on a local center rather than the global center of a 3D model. It captures the local geometry property: local radial distance distribution on view images. The view images capture both the 2D and 3D information of a 3D model.

First, we define a set of local centers by uniformly dividing the axis-aligned bounding cube of a 3D model into a set of small cubes. Since we normalize a 3D model into a bounding sphere with a radius of 1, the bounding cube is,

$$\{(x_i, y_i, z_i) | x_i, y_i, z_i \in [-1,1]\}. \quad (3)$$

If we divide the bounding cube into $N \times N \times N$ small cubes, then the coordinates of the local centers of these small cubes are,

$$\{(x_i, y_i, z_i) | x_i, y_i, z_i \in \left\{\frac{2i-N-1}{N} \middle| 1 \leq i \leq N\right\}. \quad (4)$$

One example ($N=3$, 27 local centers) is shown in Figure 2.

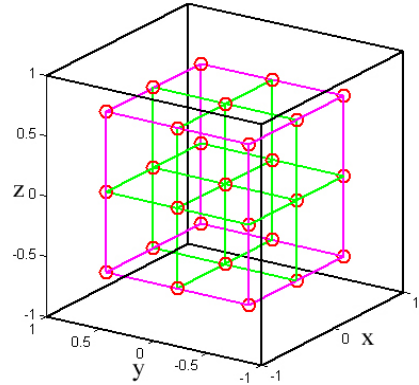

Fig. 2: Sampling local centers based on uniformly dividing the bounding cube into 27 (3×3×3) local small cubes. The circles represent the local centers. The lines between the circles are used to show their relative positions.

Then, for every vertex on the surface of the model, we find the closest local center and compute the distance $d$ between the vertex and its closest local center. We use $d$ to represent the local radial distance feature of a vertex.

Next, we encode this local feature as grayscale information into a rendered view image by color coding. We sparsely sample only 13 views for each model by setting cameras at the following locations on a cube: {(1,1,1), (-1,1,1), (-1,-1,1), (1,-1,1), (1,0,0), (0,1,0), (0,0,1), (1,0,-1), (0,1,-1), (1,1,0), (0,1,1), (1,0,1), (1,-1,0)}. They are composed of 4 top corners, 3 adjacent face centers and 6 middle edge points, respectively. In the rendering, for every vertex, we use its local radial distance $d$ as its ($r$, $g$, $b$) color values. By adopting the smooth shading, we approximate the local radial distances of the points in each face of the model. One example of the 13 rendered local radial distance feature views of a 3D model is shown in Figure 3. These rendered views capture the 2D contour information as well as the local geometry information of the 3D model.

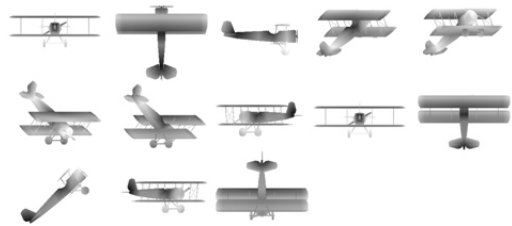

Fig. 3: 13 local radial distance feature views. The grayscale depicts the local radial distance and darker means smaller.

Finally, we compute the Zernike moments [13] (up to 10[th] order, total of 36 moments) of each view image. Therefore, the local radial distance feature of a 3D model is a 13×36 matrix.

### 4. 3D MODEL RETRIEVAL ALGORITHM

We focus on retrieval using 3D models as queries. Given a query model and a 3D model database, we propose a 3D model retrieval algorithm as follows.

(1) **Coordinates normalization.** To achieve translation and scale invariance, we first compute the bounding sphere [28, 7] of a 3D model, then translate the center of the sphere to the origin of the coordinate system and finally scale the model to make the radius of the sphere to be 1.0.

(2) **Pose normalization.** To improve retrieval accuracy, we need to align the 3D models. To concentrate on testing the performance of the hybrid descriptor itself, we manually front-pose aligned the models.

(3) **Compute the global and local radial distance shape descriptors (Sections 3.2 and 3.3).** For every model, we compute an $m \times n$ (e.g. 6×12) distance distribution matrix $G$ for the global radial distance feature and an $s \times t$ (e.g. 13×36) Zernike moments matrix $L$ for the local radial distance feature.

(4) **Compute the shape distance vector and ranking.** After comparing the performance of different distance metrics [15], such as city block distance (L1), Euclidean distance (L2), Canberra distance, correlation distance and divergence distance, we choose the L1 distance to measure the difference between two global radial distance shape descriptors ($d_g$) as well as the distance between two local radial distance shape descriptors ($d_l$),

$$d_g = \frac{1}{m \times n}\sum_{p=1}^{m}\sum_{q=1}^{n}|G(p,q) - G(p,q)|, \quad (5)$$

$$d_l = \frac{1}{s \times t}\sum_{p=1}^{s}\sum_{q=1}^{t}|L(p,q) - L(p,q)|. \quad (6)$$

Then, we normalize the two distance $d_g$ and $d_l$ by their respective maximal values over all models, and get the normalized distance $\widetilde{d_g}$ and $\widetilde{d_l}$. Finally, we add these two normalized distances together to form the hybrid shape distance $d$,

$$d = \widetilde{d_g} + \widetilde{d_l}. \quad (7)$$

After computing the distances between the query model and every model in the database, we get a shape distance vector $D = \{d_1, d_2, \cdots, d_T\}$, where $T$ is the total number of models in the database. Finally, we sort $D$ in ascending order to give the retrieval results for the query model.

## 5. EXPERIMENTS AND DISCUSSIONS

We tested our global and local radial distances based hybrid descriptor (Hybrid, and its global and local components are referred as Global and Local) on the Princeton Shape Benchmark Database (PSB) [24] and the National Taiwan University Benchmark (NTU) [4]. We also compared it with two typical histogram-based descriptors which are shape distribution (D2) [21, 22] and shape histogram (SHELL, SECTOR and SPIDER) [1], and one famous and typical view-based approach named LightField descriptor (LF) [4].

After getting the distance matrix for a set of models, we mainly used the tools provided in PSB to generate the precision-recall plots [24]. Precision-recall plot is one of the most often used measures to evaluate the performance of a retrieval algorithm. Recall means how much percentage of a class has been retrieved among the top $K$ list while precision indicates how much percentage of the top $K$ models belongs to the same type as the query model.

### 5.1 Princeton Database

The test dataset of Princeton Shape Benchmark Database (PSB) [24] contains 907 models, classified into 131 classes. For the precision-recall plots of D2 and SHELL (shape histogram) descriptors, we refer to the experiment results in [24] and these two descriptors are rotation-invariant. For the SECTOR and SPIDER descriptors, we plot the results [24] of their original algorithms as a reference and they used the PCA [11] to align the models. For the LightField descriptor, we generate the precision-recall plot by running their program on the front-pose aligned PSB database. Figure 4 shows the precision-recall plots of the above methods and our hybrid descriptor (together with its two components Global and Local).

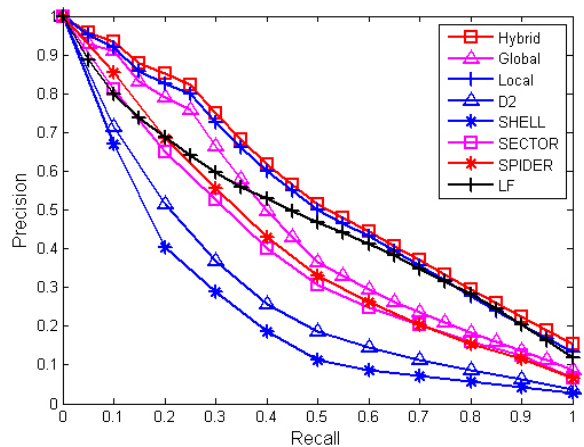

Fig. 4: PSB database: precision-recall plots.

If we measure the average precision using Area Under Curve (AUC), the Average Precision (AP) of the six descriptors are: Hybrid: 62.84%; D2: 22.25%, SHELL: 19.61%, SECTOR: 27.15%, SPIDER: 27.96%, LF: 55.99%.

As can be seen from Figure 4 and their AP values, we can conclude that our hybrid descriptor outperforms D2, shape histogram (SHELL, SECTOR and SPIDER) and LF descriptors. It is also indicated that in the case of PSB database, the local radial distance performs better than the global radial distance and it contributes more to the superior performance of the hybrid shape descriptor. Though the global radial distance performs not better than LF, it still contributes to the overall performance of the hybrid shape descriptor. It takes on average only 0.07 second for a PC with an Intel Core 2 2.66GHz CPU to compute the global radial distance feature of a 3D model in the PSB database. Therefore, it does not add much computation.

For the 131 classes in the PSB database, we found that our hybrid descriptor performs better in retrieving classes such as computer monitor, chair (either desk or dinning), door, electrical guitar, eyeglass, fish, hand gun, human, hourglass, race car, rabbit, sedan and sword, and has inferior performance in retrieving barn, bush, butterfly, cabinet, satellite, staircase, two-story home and vase models. The performances of other classes are just fall in-between. Figure 5 gives several typical classes' precision-recall plots of our hybrid descriptor and the LightField descriptor.

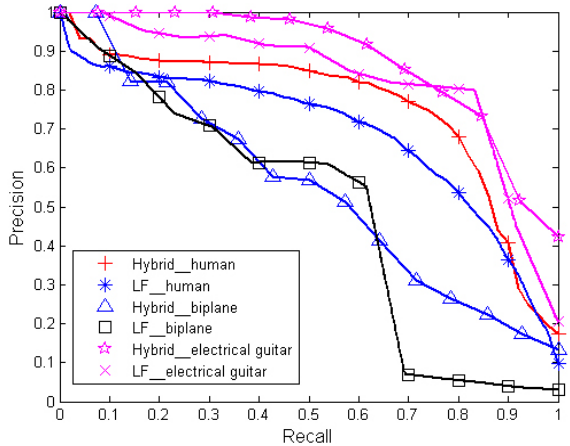

Fig. 5: PSB database class precision: biplane, human and electrical guitar.

## 5.2 NTU Database

The NTU 3D Model Benchmark [4] contains 1833 3D models. 549 3D models are clustered into 47 classes and the rest 1284 models are classified as "miscellaneous". We tested the performance using the classified 549 models. Figure 6 plots the precision-recall plots of our hybrid descriptor and the LightField descriptor.

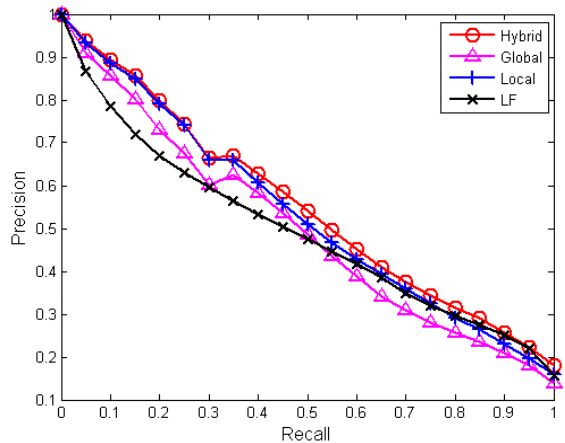

Fig. 6: NTU database: precision-recall plots.

As can be seen, the performance of our hybrid shape descriptor is still better than the LightField descriptor. Their Average Precision (AP) values are: Hybrid: 62.41%, LF: 56.53%. For this database, the global radial distance can achieve a comparable performance as that of the local radial distance. However, neither of them can outperform LF for all the recall values. When we combine them to form the hybrid distance, it exceeds LF along all the path of the precision-recall curve. We can also find that the local radial distance has more contribution to the performance of the hybrid shape descriptor when recall is larger than 0.3, which means retrieving more than 30% relevant models. This suggests adding the global radial distance can push the relevant model forward in the retrieval list. From both PSB and NTU database, we can see that the precision-recall curves of the hybrid descriptor are always above the curves of either the local or the global distance descriptors.

For the classified 47 classes, our hybrid shape descriptor does well in retrieving the following classes: ball, bike, book, bottle, car, casket, chip, door, driver, facemask, glass, guitar, gun, hat, hydrant, knife, motorcycle, plane, shield, starship, stick, submarine and sword. It has low performance for drum and pot classes. Others' performances fall in-between. Figure 7 shows the first 10-nearest neighbor results for two typical retrieval examples. We can see that our method can achieve more accurate retrieval lists for these types of models.

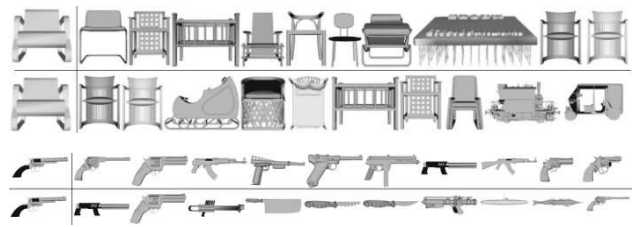

Fig. 7: Two retrieval examples: our hybrid descriptor: the first row; LF: the second row. The left most model of each row is the query model.

## 6. CONCLUSION AND FUTURE WORK

We have proposed a hybrid shape descriptor by combining the histogram-based approach and view-based approach. It integrates their advantages and simultaneously avoids their shortcomings. The hybrid shape descriptor is comprised of a global radial distance feature and a local radial distance feature. We can efficiently extract the global radial distance feature using a histogram-based approach and compute an effective local feature based on a view-based approach. The local feature already can outperform the well-known view-based approach, LightField [4]. Adding the global feature can further improve the performance without adding much computation load. To compute the global radial distance, we compute the distance from the center of the bounding sphere and sample sparsely only on the face centers by area-weighting. We define the local radial distance feature by dividing the axis-aligned bounding cube of a 3D model into a set of small cubes and compute the local radial distance based on the closest local center. To compute the difference between the local radial distance features, we encode them as color information into 13 view images of the model and then use the Zernike moments to compare two views.

We have presented a hybrid radial distance feature based retrieval algorithm. The hybrid feature improves the retrieval performance and can achieve better results compared to several typical histogram-based approaches like shape histogram [1], shape distribution [21, 22] and the well-know view-based approach, LightField [4].

It should be noted that this paper has been primarily concerned with constructing an effective shape descriptor. Therefore, testing the influence of different alignment algorithms, such as PCA and continuous PCA (CPCA) [27] on our descriptor is our main future work.

## 7. REFERENCES

[1] Annkherst, M., Kastenmuller, G., Kriegel, H.P., and Seidl, T., "3D shape histograms for similarity search and classification in spatial databases", 6th International Symposium on Advances in Spatial Databases, pp. 207-226, 1999.

[2] Ben-Chen, M., and Gotsman, C., "Characterizing shape using conformal factors", Eurograhphics Workshop on 3D Object Retrieval, 3DOR08, 2008.

[3] Belongie, S., Malik, J., and Puzicha, J., "Shape matching and object recognition using shape contexts", IEEE Trans. on PAMI, Vol. 24, No. 24, pp. 509-522, 2002.

[4] Chen, D.Y., Tian, X.P., Shen, Y.T., and Ouhyoung, M., "On visual similarity based 3D model retrieval", Computer Graphics Forum (Eurographics'03), Vol. 22, No. 3, pp. 223-232, 2003.

[5] Elad, M., Tal A., and Ar. S., "Content based retrieval of VRML objects - an iterative and interactive approach", 6th Eurographics Workshop on Multimedia, pp. 97-108, 2001.

[6] Frome, A., Huber, D., Kolluri, R., Bülow, T., and Malik, J., "Recognizing objects in range data using regional point descriptors", 8th European Conference on Computer Vision (ECCV), 2004.

[7] Gärtner, B., "Fast and robust smallest enclosing balls", 7th Annual European Symposium on Algorithms (ESA), Lecture Notes in Computer Science, Vol. 164, pp. 325-338, 1999.

[8] Hilaga, M., Shinagawa, Y., Kohmura, T., and Kunii, T. L., "Topology matching for fully automatic similarity estimation of 3d shapes", 28th Annual Conference on Computer Graphics and Interactive Techniques, SIGGRAPH'01, pp. 203-212, 2001.

[9] Huang, K.S., and Trivedi, M.M.: "3D shape context based gesture analysis integrated with tracking using omni video array", IEEE Conference on Computer Vision and Pattern Recognition (CVPR), 2005.

[10] Jeannin, S., Cieplinski, L., Ohm, J.R., and Kim, M., "MPEG-7 visual part of eXperimentation model version 7.0", ISO/IEC JTC1/SC29/WG11/N3521, 2000.

[11] Jolliffe, I.T., "Principal component analysis", 2nd edition, Springer, 2002.

[12] Kazhdan, M., Funkhouser, T., and Rusinkiewicz, S., "Rotation invariant spherical harmonic representation of 3D shape descriptors", Symposium on Geometry Processing, pp. 156-164, 2003.

[13] Khotanzad, A., and Hong, Y., "Invariant image recognition by Zernike moments", IEEE Trans. on PAMI, Vol. 12, No. 5, pp. 204-241, 1990.

[14] Körtgen, M., Park, G.J, Novotn, M., and Klein, R., "3D shape matching with 3D shape contexts", 7th Central European Seminar on Computer Graphics (CESCG), 2003.

[15] Laga, H., and Nakajima, M., "Supervised learning of similarity measures for content-based 3D model retrieval", 3rd International Conference on Large-Scale Knowledge Resources (LKR), Lecture Notes in Computer Science, pp. 210-225, 2008.

[16] Laga, H., Takahashi, H., and Nakajima, M.,"Spherical wavelet descriptors for content based 3D model retrieval", International Workshop on Advanced Image Technology (IWAIT), 2006.

[17] Lowe, D.G., "Distinctive image features from scale-invariant keypoints", International Journal of Computer Vision, Vol. 60, No. 2, pp. 91-110. 2004.

[18] Mori, G., and Malik, J., "Recognizing objects in adversarial clutter: breaking a visual CAPTCHA", IEEE Conference on Computer Vision and Pattern Recognition (CVPR), pp. 134-141, 2003.

[19] Natraj, I., Subramaniam, J., Kuiyang, L., Yagnanarayanan K., and Karthik, R., "Three-dimensional shape searching: state-of-the-art review and future trends", Computer-Aided Design, Vol. 37, No. 5, pp. 509-530, 2005.

[20] Petros, D., and Apostolos, A., "A compact multi-view descriptor for 3D object retrieval", 7th International Workshop on Content-Based Multimedia Indexing, pp.115-119, 2009.

[21] Robert, O., Thomas, F., Bernard, C., and David, D., "Matching 3D models with shape distributions", Shape Modeling International (SMI), pp. 154-166, 2001.

[22] Robert, O., Thomas, F., Bernard, C., and David, D., "Shape distributions", ACM Trans. on Graphics, Vol. 21, No. 4, pp. 807-832, 2002.

[23] Ryutarou, O., Kunio, O., Takahiko, F., and Tomohisa, B., "Salient local visual features for shape-based 3D model retrieval", Shape Modeling International (SMI), pp. 93-102, 2008.

[24] Shilane, P., Min, P., Kazhdan, M., and Funkhouser, T., "The princeton shape benchmark", Shape Modeling International (SMI), pp. 167-178, 2004.

[25] Sundar, H., Silver, D., Gagvani, N., and Dickinson, S., "Skeleton based shape matching and retrieval", Shape Modeling International (SMI), pp. 130-139, 2003.

[26] Tangelder, J.W.H., and Veltkamp, R.C., "A survey of content based 3D shape retrieval methods", Shape Modeling International (SMI), pp.145-156, 2004.

[27] Vranic, D., "3D Model Retrieval", PhD Thesis, Leipzig University, 2004.

[28] Welzl, E., "Smallest enclosing disks (balls and ellipsoids)", New Results and New Trends in Computer Science, Lecture Notes in Computer Science, Vol. 555, pp. 359-370, 1991.

[29] Zhang, D. and Luo, G., "An integrated approach to shape based image retrieval", 5th Asian Conference on Computer Vision (ACCV), pp. 652-657, 2002.